\documentclass{elsart}
\usepackage{amssymb}
\usepackage{epsfig}
\usepackage{graphics}
\usepackage{color}
\definecolor{mygreen}{rgb}{0,.8,0}

\def\ifm#1{\relax\ifmmode#1\else$#1$\fi}  
    \def\x{\ifm{\times}}

  \def\dif{\hbox{d\kern.5mm}}

\let\cal=\mathcal   \def\ORD#1!{\ifm{{\cal O}\hbox{(#1)}}}
\def\pt#1,#2,{\ifm{#1\x10^{#2}}}

\def\ifm#1{\relax\ifmmode#1\else$#1$\fi}

  \def\x{\ifm{\times}}
  
\def\pt#1,#2,{\ifm{#1\x10^{#2}}}  \def\dif{\ifm{{\rm d}\,}}

\makeatletter
\newdimen\z@ \z@=0pt 
\newskip\z@skip \z@skip=0pt plus0pt minus0pt
\def\m@th{\mathsurround=\z@}
\def\ialign{\everycr{}\tabskip\z@skip\halign} 
\def\eqalign#1{\null\,\vcenter{\openup\jot\m@th
  \ialign{\strut\hfil$\displaystyle{##}$&$\displaystyle{{}##}$\hfil
      \crcr#1\crcr}}\,}
\makeatother

\newcommand{\aff}[2]{Dipartimento di Fisica dell'Universit\`a #1 e Sezione INFN, #2, Italy.}
\newcommand{\affd}[1]{Dipartimento di Fisica dell'Universit\`a e Sezione INFN, #1, Italy.}
\begin{document}

\begin{frontmatter}

\title{
Study of the decay $\phi\rightarrow$f$_0$(980)$\gamma
\rightarrow\pi^+\pi^-\gamma$ with the KLOE detector
}
\collab{The KLOE Collaboration}

\author[Na]{F.~Ambrosino},
\author[Frascati]{A.~Antonelli},
\author[Frascati]{M.~Antonelli},
\author[Roma3]{C.~Bacci},
\author[Karlsruhe]{P.~Beltrame},
\author[Frascati]{G.~Bencivenni},
\author[Frascati]{S.~Bertolucci},
\author[Roma1]{C.~Bini},
\author[Frascati]{C.~Bloise},
\author[Roma1]{V.~Bocci},
\author[Frascati]{F.~Bossi},
\author[Frascati,Virginia]{D.~Bowring},
\author[Roma3]{P.~Branchini},
\author[Roma1]{R.~Caloi},
\author[Frascati]{P.~Campana},
\author[Frascati]{G.~Capon},
\author[Na]{T.~Capussela},
\author[Roma3]{F.~Ceradini},
\author[Frascati]{S.~Chi},
\author[Na]{G.~Chiefari},
\author[Frascati]{P.~Ciambrone},
\author[Virginia]{S.~Conetti},
\author[Frascati]{E.~De~Lucia},
\author[Roma1]{A.~De~Santis},
\author[Frascati]{P.~De~Simone},
\author[Roma1]{G.~De~Zorzi},
\author[Frascati]{S.~Dell'Agnello},
\author[Karlsruhe]{A.~Denig},
\author[Roma1]{A.~Di~Domenico},
\author[Na]{C.~Di~Donato},
\author[Pisa]{S.~Di~Falco},
\author[Roma3]{B.~Di~Micco},
\author[Na]{A.~Doria},
\author[Frascati]{M.~Dreucci},
\author[Frascati]{G.~Felici},
\author[Frascati]{A.~Ferrari},
\author[Frascati]{M.~L.~Ferrer},
\author[Frascati]{G.~Finocchiaro},
\author[Roma1]{S.~Fiore},
\author[Frascati]{C.~Forti},
\author[Roma1]{P.~Franzini},
\author[Frascati]{C.~Gatti},
\author[Roma1]{P.~Gauzzi},
\author[Frascati]{S.~Giovannella},
\author[Lecce]{E.~Gorini},
\author[Roma3]{E.~Graziani},
\author[Pisa]{M.~Incagli},
\author[Karlsruhe]{W.~Kluge},
\author[Moscow]{V.~Kulikov},
\author[Roma1]{F.~Lacava},
\author[Frascati]{G.~Lanfranchi},
\author[Frascati,StonyBrook]{J.~Lee-Franzini},
\author[Karlsruhe]{D.~Leone},
\author[Frascati]{M.~Martini},
\author[Na]{P.~Massarotti},
\author[Frascati]{W.~Mei},
\author[Na]{S.~Meola},
\author[Frascati]{S.~Miscetti},
\author[Frascati]{M.~Moulson},
\author[Karlsruhe]{S.~M\"uller},
\author[Frascati]{F.~Murtas},
\author[Na]{M.~Napolitano},
\author[Roma3]{F.~Nguyen},
\author[Frascati]{M.~Palutan},
\author[Roma1]{E.~Pasqualucci},
\author[Roma3]{A.~Passeri},
\author[Frascati,Energ]{V.~Patera},
\author[Na]{F.~Perfetto},
\author[Roma1]{L.~Pontecorvo},
\author[Lecce]{M.~Primavera},
\author[Frascati]{P.~Santangelo},
\author[Roma2]{E.~Santovetti},
\author[Na]{G.~Saracino},
\author[Frascati]{B.~Sciascia},
\author[Frascati,Energ]{A.~Sciubba},
\author[Pisa]{F.~Scuri},
\author[Frascati]{I.~Sfiligoi},
\author[Frascati]{T.~Spadaro},
\author[Roma1]{M.~Testa},
\author[Roma3]{L.~Tortora},
\author[Roma1]{P.~Valente},
\author[Karlsruhe]{B.~Valeriani},
\author[Frascati]{G.~Venanzoni},
\author[Roma1]{S.~Veneziano},
\author[Lecce]{A.~Ventura},
\author[Roma1]{S.~Ventura},
\author[Karlsruhe]{R.~Versaci},
\author[Beijing,Frascati]{G.~Xu},

\address[Beijing]{Institute of High Energy 
Physics of Academica Sinica, Beijing, China.}
\address[Frascati]{Laboratori Nazionali di Frascati dell'INFN, 
Frascati, Italy.}
\address[Karlsruhe]{Institut f\"ur Experimentelle Kernphysik, 
Universit\"at Karlsruhe, Germany.}
\address[Lecce]{\affd{Lecce}}
\address[Moscow]{Institute for Theoretical 
and Experimental Physics, Moscow, Russia.}
\address[Na]{Dipartimento di Scienze Fisiche dell'Universit\`a 
``Federico II'' e Sezione INFN,
Napoli, Italy}
\address[Pisa]{\affd{Pisa}}
\address[Energ]{Dipartimento di Energetica dell'Universit\`a 
``La Sapienza'', Roma, Italy.}
\address[Roma1]{\aff{``La Sapienza''}{Roma}}
\address[Roma2]{\aff{``Tor Vergata''}{Roma}}
\address[Roma3]{\aff{``Roma Tre''}{Roma}}
\address[StonyBrook]{Physics Department, State University of New 
York at Stony Brook, USA.}
\address[Virginia]{Physics Department, University of Virginia, USA.}
{\small Corresponding author: Cesare Bini, e-mail
  cesare.bini@roma1.infn.it, tel +390649914266, fax +39064957697}
\begin{abstract}
  We measured, with the KLOE detector, the spectrum of 
  $\pi^+\pi^-$ invariant mass in a sample of $6.7\times 10^{5}$ 
  $e^+e^-\rightarrow\pi^+\pi^-\gamma$ events with the photon at large polar
  angle ($\theta_{\gamma}>45^{\circ}$)
  at a centre of mass energy $\sqrt{s}$ around the $\phi$ mass.
  We observe in this spectrum a clear contribution from the intermediate process
  $\phi\rightarrow~\mbox{f}_{0}(\mbox{980})\gamma$.
  A sizeable effect is also observed in the distribution of the pion
  forward-backward 
  asymmetry. We use different theoretical models to fit the spectrum and we
  determine the $\mbox{f}_{0}$
  mass and coupling constants to the $\phi$, to $\pi^{+}\pi^{-}$ and
  to $K\bar{K}$. 
\end{abstract}
\end{frontmatter}
\section{Introduction}
\label{uno}
The $\phi$(1020) radiative decays to f$_0$(980) and a$_0$(980) 
play an important role in the investigation of 
the controversial structure of the lighter scalar 
mesons~\cite{AchasovKL,motivations}. 
At KLOE, we detect the f$_0$ through its
decay to $\pi\pi$ via the decay chain
$e^+e^-\rightarrow\phi\rightarrow\mbox{f}_{0}
\rightarrow\pi\pi\gamma$.
KLOE has already published studies on $\phi$ decays to 
f$_0\gamma$ and a$_0\gamma$ looking for
the final states $\pi^0\pi^0\gamma$~\cite{ppg2002} and 
$\eta\pi^0\gamma$~\cite{epg2002}, respectively. 
On the contrary, the decay chain $e^+e^-\rightarrow\phi\rightarrow\mbox{f}_{0}(\mbox{a}_{0})
\rightarrow K\bar{K}\gamma$ is kinematically
suppressed, and has not been observed yet.      
In this paper we present a study of the f$_0$ decay to $\pi^+\pi^-$, based on an
integrated luminosity of 350 pb$^{-1}$ collected at the collider DA$\Phi$NE
during the years 2001 and 2002, at a centre of mass energy $\sqrt{s}$ around
the $\phi$ mass $m_{\phi}= 
1019.45~\mbox{MeV}$
within $\pm 0.5$ MeV (``on-peak'' data). 
The only previous search for this decay has been published by the CMD-2 
collaboration~\cite{CMD2}, mainly based on an energy scan around the $\phi$
mass peak.

We look for f$_0\rightarrow\pi^+\pi^-$ decays
in events $e^+e^-\rightarrow\pi^{+}\pi^{-}\gamma$.
Only a small fraction of the $\pi^+\pi^-\gamma$ events originates
from the radiative decay $\phi\rightarrow$f$_0\gamma$ with
f$_0\rightarrow\pi^+\pi^-$. The main contribution
is given by $e^+e^-\rightarrow\pi^{+}\pi^{-}\gamma$ events with
a photon from initial state (ISR) or final state (FSR) radiation.
The amplitude of each contribution is characterised by a 
different spectrum of the $\pi^+\pi^-$ invariant mass $m$, 
and of the photon polar angle $\theta_{\gamma}$ measured 
with respect to the beam axis. 
In particular, the ISR is the dominant contribution
for small photon polar angles, allowing to extract the
$e^+e^-\rightarrow\pi^+\pi^-$ cross-section below the $\phi$ mass with the
so called radiative return method \cite{hadronic};
at large
values of $\theta_{\gamma}$ the ISR contribution is strongly reduced, so that
the other processes can be observed in this region only.
A smaller contribution comes from the decay
$\phi\rightarrow\rho^{\pm}\pi^{\mp}$ with 
$\rho^{\pm}\rightarrow\pi^{\pm}\gamma$ (we call it $\rho\pi$ term in the
following).
It contributes in the low mass region,
$400<m<600~\mbox{MeV}$, with total branching ratio 
BR($\phi\rightarrow\rho^{\pm}\pi^{\mp}$)$\times$
BR($\rho^{\pm}\rightarrow\pi^{\pm
}\gamma$)$\sim4\times 10^{-5}$. Finally, the possibility to observe the decay
chain $\phi\rightarrow$f$_0$(600)$\gamma\rightarrow\pi^+\pi^-\gamma$ is 
considered.  

The $\pi^+\pi^-$ pair has different quantum numbers
whether it is produced through FSR and f$_0$ decay
or ISR: J$^{PC}$=0$^{++}$ in the former case, and
J$^{PC}$=1$^{--}$ in the latter. A sizeable interference term between 
FSR and f$_0$ decay is expected in the $m$ spectrum. On the other hand 
any interference term between two amplitudes of opposite
charge conjugation gives rise to $C$-odd terms that change
sign by the interchange of the two pions.  Therefore, 
the interference between ISR and FSR or f$_0$ decay, results in a null
contribution in the $m$ spectrum 
for symmetric cuts on $\theta_{\gamma}$ and $\theta_{\pi^{\pm}}$,
and in a sizable forward-backward asymmetry, $A_c$, defined as:\\
\begin{equation}
A_{c}={{N(\theta_{\pi^+}>90^{\circ})-N(\theta_{\pi^+}<90^{\circ})}\over
 {N(\theta_{\pi^+}>90^{\circ})+N(\theta_{\pi^+}<90^{\circ})}},
\label{asymmetry}
\end{equation}
where the angle $\theta_{\pi^+}$ is defined with respect to the direction
of the incoming electron beam. 
\section{Experimental set-up}
\label{due}
DA$\Phi$NE is an $e^+e^-$-collider with a peak luminosity of about 
10$^{32}$cm$^{-2}$s$^{-1}$ at a centre of mass
energy $\sqrt{s}=m_{\phi}=1.02$ GeV. The beams collide with a crossing angle of
$\pi$-0.025 rad. The KLOE detector consists of
a large-volume cylindrical drift chamber~\cite{dch} (3.3~m length
and 2~m radius), operated with a 90\% helium-10\% isobutane gas mixture,
surrounded by a sampling calorimeter~\cite{calo} made of lead and
scintillating fibres providing a solid angle coverage of 98\%. 
The tracking chamber and the calorimeter are surrounded by 
a superconducting coil 
that produces a solenoidal field $B$=0.52~T.
The drift chamber has a momentum resolution of $\sigma(p_\perp)/p_\perp \sim $
0.4\%.  Photon energies and arrival times are measured by the calorimeter with
resolutions of $\sigma_E/E=5.7\%/\sqrt{E({\rm GeV})}$ and 
$\sigma_t=54{\rm ps}/\sqrt{E({\rm GeV})}\oplus 50$ ps.
The trigger~\cite{TRGnim} is based on the detection of at least
two energy deposits in the
calorimeter above a threshold that ranges between 50 and 150~MeV. 
The trigger includes a cosmic ray veto based on
large energy deposits in the outermost calorimeter layers.
\section{Event selection}
\label{tre}
We select $\pi^+\pi^-\gamma$ events by requiring a
reconstructed vertex close to the interaction region 
with two tracks of opposite charge, 
emitted with polar angles above 
$45^\circ$ ($\theta_{\pi^{\pm}}>45^{\circ}$). 
We suppress the ISR component by requiring the polar angle of the
total missing momentum to be larger than 
$45^\circ$ ($\theta_{\gamma}>45^{\circ}$).
Both tracks are extrapolated to the calorimeter.
A likelihood variable, based on the time of flight 
and on the shower profile (see ref.~\cite{hadronic}),
is used to select pions. A cut on this variable reduces 
the background due to $e^+e^-\gamma$ events to a negligible level.\\
In order to remove $\pi^+\pi^-\pi^0$ and $\mu^+\mu^-\gamma$
events, we define the track mass variable $M_T$ 
as the solution of the equation:
\begin{equation}
  |\vec{p}_{\phi}-\vec{p}_1-\vec{p}_2|=
  E_{\phi}-\sqrt{p_1^2+M_T^2}-\sqrt{p_2^2+M_T^2},
  \label{trackmass}
\end{equation}
where $\vec{p}_1$ and $\vec{p}_2$ are the momenta of the two tracks, and 
where $E_{\phi}$ and $\vec{p}_{\phi}$ are the $\phi$ energy and 
momentum, respectively. These are evaluated run by run using samples of 
Bhabha scattering events. Eq.~(\ref{trackmass}) 
is verified by events with two particles of mass $M_T$
and a third particle with null mass. 
$M_T$ is required to be equal to the pion
mass within $\pm$10 MeV .
To reduce the residual $\pi^+\pi^-\pi^0$ contamination 
and to remove badly reconstructed
$e^+e^-\rightarrow\pi^+\pi^-$ events with soft photon emitted, we require a
calorimeter cluster matching the missing energy and momentum. 
The cluster is required to be non associated to tracks, to have 
an energy $E_{\gamma}>10~\mbox{MeV}$, and to have a
time compatible with a photon coming from the interaction vertex, 
$|t_{\gamma}-R_{\gamma}/c|<5\sigma_t(E_{\gamma})$, where $t_{\gamma}$ is the
cluster time, $R_{\gamma}$ is the flight distance, and $\sigma_t(E_{\gamma})$
is the time resolution for photons of energy $E_{\gamma}$.  
The
requirement $E_{\gamma}>10~\mbox{MeV}$ translates in an effective cut 
$m<1009~\mbox{MeV}$.
Finally, the angle $\Omega$ between the missing momentum and 
the photon direction derived from the cluster position,
has to be below 0.03+$3/E_{\gamma}(\mbox{MeV})$~rad. 
The dependence of the $\Omega$ cut on $E_{\gamma}$
reflects that of the cluster position resolution 
on the photon energy. \\

\begin{figure}[ht]
  \centering
  \epsfig{file=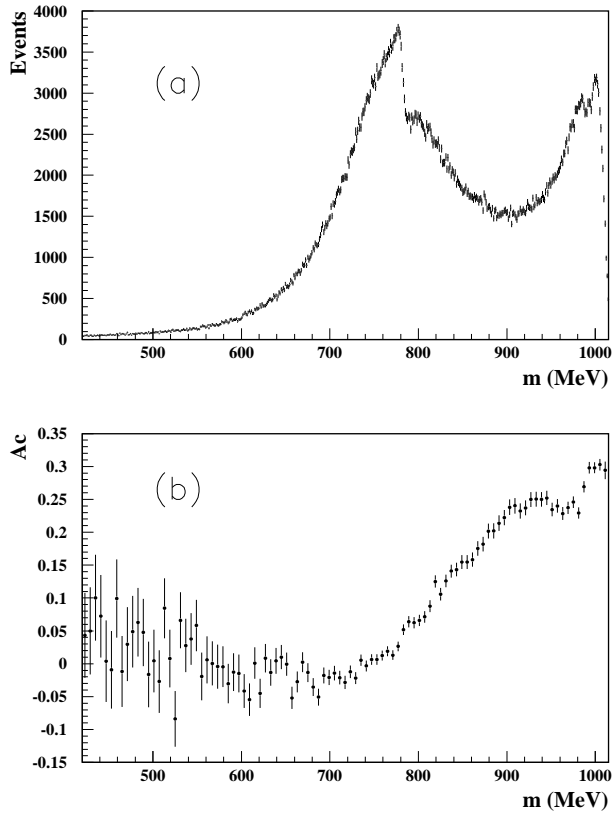,width=9cm}
  \caption{\small{\it (a) $\pi^+\pi^-$ invariant mass spectrum of the selected
      sample. The spectrum is dominated by the ISR component, showing  
      the $\rho$-$\omega$ interference
      pattern.  
      The signal of the f$_0$(980) appears as a small peak around 980~MeV. 
      The drop for $m>1000~\mbox{MeV}$ is due to the 
      drop of the detection efficiency for low energy photons.
      (b) Forward-backward asymmetry defined in Eq.~(\ref{asymmetry}) as a function
      of $m$. The dip in the region of the f$_0$(980) is evident.
  }}
  \label{dati}
\end{figure}

We select 6.7$\times 10^{5}$ events. The spectrum of the $\pi\pi$ invariant mass $m$ 
for these events and the forward-backward asymmetry dependence on $m$ are shown in
Fig.~\ref{dati}. 
We observe a small bump in the $m$ spectrum in the region 
where the f$_0$(980) is expected to be.
A signal is observed also as a dip in the forward-backward asymmetry $A_{c}$, 
for the same values of $m$.\\ 
\begin{figure}[ht]
  \centering
  \epsfig{file=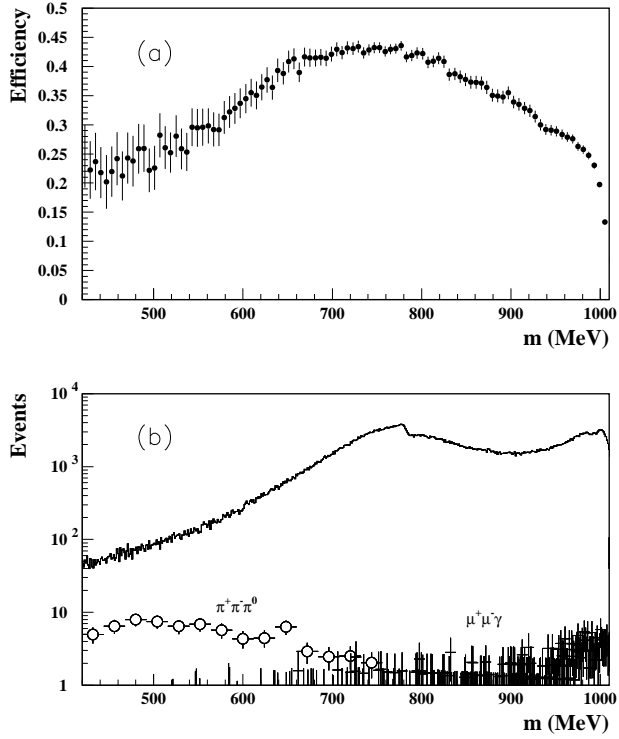,width=9cm}
  \caption{\small{\it (a) Total efficiency as a function of $m$. (b)
      The Monte Carlo expected contributions of the main background
      sources, $\pi^+\pi^-\pi^0$ (open circles) and $\mu^+\mu^-\gamma$
      (crosses),  
      normalised to the 
      integrated luminosity, and compared to the data spectrum. 
  }}
  \label{effic}
\end{figure}

Total efficiency and residual background distributions are shown in Fig.~\ref{effic},
as evaluated by Monte Carlo with corrections based on data control samples
\cite{nota1}. The simulation of the ISR and FSR contributions is based on the
EVA generator \cite{binner}. 
The efficiency decrease at low masses is due to the increased 
occurrence of low-p$_T$ pions with $\theta_{\pi^{\pm}}<45^{\circ}$ that escape the
selection; the decrease for higher masses, starting from $\sim$ 800~MeV, 
is partly due to the cosmic ray veto and partly to the photon detection efficiency. 
In fact, high momentum pions, which deposit large energy in the outermost 
calorimeter layers, veto the event with high probability. Moreover,
low energy photons ($E_{\gamma}<20~\mbox{MeV}$) are detected with an
efficiency lower than 80\%.  
The efficiency for the cosmic veto is evaluated using samples of
pre-scaled events with no veto applied. The photon detection efficiency 
is measured as a function of $E_{\gamma}$ from $\pi^+\pi^-\pi^0$ and 
$e^+e^-\gamma$ control samples~\cite{calo}.\\
After the selection, $\phi\rightarrow\pi^+\pi^-\pi^0$ decays  
give the only significant contribution to the background.
\section{Description of the fit}
\label{quattro}
We fit the $\pi^{+}\pi^{-}$ invariant mass spectrum,
$dN/dm$, with the function:
\begin{eqnarray}
  \nonumber
      {{dN}\over{dm}}&=&L_{int}\epsilon(m)\left({{d\sigma_{ISR}}\over{dm}}+
      {{d\sigma_{FSR}}\over{dm}}+{{d\sigma_{\rho\pi}}\over{dm}}+
      {{d\sigma_{\rm scal}}\over{dm}}\pm
      {{d\sigma^{INT}_{\rm scal+FSR}}\over{dm}}\right)+
      \\
      & & {back}, 
      \label{total}
\end{eqnarray}
where 
$L_{int}$ is the integrated luminosity,
$\epsilon(m)$ is the selection efficiency, 
and ${back}$ is the residual background.
The first three terms in parenthesis are here called the ``non-scalar'' terms.
The analytic expressions for the first and second terms, ISR and FSR, 
are taken from ref.~\cite{Achasov},
while the $\rho\pi$ term is taken from ref.~\cite{Achasovrpi}. 
The pion form factor~\cite{Santamaria},
entering the ISR term, depends on the masses and widths 
of the $\rho^0$, $\omega$ and $\rho'$ mesons, and on the two 
non dimensional parameters $\alpha$ and $\beta$, which correspond to the sizes of 
the $\omega$ and $\rho'$ contributions, respectively. 
We leave the quantities $m_{\rho^0}$, $\Gamma_{\rho^0}$, $\alpha$, and $\beta$
as free parameters of the fit while
the masses and the widths of the $\omega$ and $\rho'$ mesons are 
fixed to the PDG values~\cite{PDG}.
The $\rho\pi$ term is multiplied by a scale factor, a$_{\rho\pi}$, 
which is expected to be equal to unity. If a$_{\rho\pi}$=1 the number of
$\rho\pi$ events corresponds to approximately 1\% of the total, broadly distributed in
the low mass region.  
The possible interference between the $\rho\pi$ and the scalar terms is neglected.
The last two terms in parenthesis, scal and scal$+$FSR, 
depend on the amplitude 
for the decay $\phi\rightarrow$f$_0\gamma\rightarrow\pi^+\pi^-\gamma$, the
latter being the interference term between f$_0$ and FSR.
The scal$+$FSR term
can be either added (constructive interference) or subtracted (destructive
interference). \\

We perform three fits corresponding to three different approaches in the
description of the scalar amplitude. \\
The first fit is the Kaon-Loop fit (KL)~\cite{AchasovKL,Achasov}: 
the $\phi$ couples to the scalar through a loop of charged kaons. The
formalism allows the inclusion of more than one scalar meson.
For each scalar meson there are three free parameters of the fit: 
the mass and
the couplings to $K^+K^-$ and to $\pi^+\pi^-$. 
For the f$_0$ scalar meson only, the amplitude reduces to:
\begin{equation}
A_{KL}=g(m^2)e^{i\delta(m)}{{g_{{\rm f_0}K^+K^+}
g_{{\rm f_0}\pi^+\pi^-}}\over{(s-m^2)D'_{\rm f_0}(m)}},
\label{KL}
\end{equation}
where $s$ is the square of the centre of mass energy, $g_{{\rm f_0}K^+K^+}$
and $g_{{\rm f_0}\pi^+\pi^-}$ are the two couplings,
$g(m^2)$ is the kaon-loop function~\cite{AchasovKL}, 
$\delta(m)$ is the phase of the background to the $\pi\pi$ elastic scattering, 
and $D'_{\rm f_0}$ is the f$_0$ inverse propagator with the finite width 
corrections~\cite{AchasovKL}. \\
In the second fit, called No-Structure fit (NS)~\cite{IsiMai}, 
a direct coupling $g_{\phi {\rm f_0}\gamma}$ of the 
$\phi$ to the f$_0$ is assumed, with a subsequent coupling 
$g_{{\rm f_0}\pi^+\pi^-}$ of the f$_0$ to the $\pi^+\pi^-$ pair.
The f$_0$ amplitude is a Breit-Wigner with a mass dependent
width~\cite{flatte} added to a polynomial complex function, the continuum, 
to allow an appropriate dumping of the resulting line shape. 
The amplitude depends on eight parameters: 
the mass $m_{\rm f_0}$; the three couplings 
$g_{\phi{\rm f_0}\gamma}$, $g_{{\rm f_0}\pi^+\pi^-}$,
and $g_{{\rm f_0}K^+K^-}$; four parameters describing the continuum: 
two coefficients $a_0$ and $a_1$ and two phases $b_0$ and $b_1$. 
The amplitude is:
\begin{equation}
  A_{NS}={{g_{\phi{\rm f_0}\gamma} g_{{\rm f_0}\pi^+\pi^-}}
    \over{D_{\rm f_0}(m)}}+{{a_0}\over{m_{\phi}^2}}e^{ib_0p_{\pi}(m)}+
  a_1{{m^2-m_{\rm f_0}^2}\over{m_{\phi}^4}}e^{ib_1p_{\pi}(m)},
\label{NS}
\end{equation}
where $p_{\pi}(m)=\sqrt{m^2/4-m^2_{\pi}}$ is the $\pi$ momentum in the f$_0$ rest
frame and  
$D_{\rm f_0}(m)$ is the f$_0$ inverse propagator:
\begin{eqnarray}
  \nonumber
  D_{\rm f_0}(m)&=&m^2-m^2_{\rm f_0}+
  i\left({{g^2_{{\rm f_0}\pi\pi}}\over{16\pi}}\sqrt{1-{{4m^2_{\pi}}\over{m^2}}}+
  {{g^2_{{\rm f_0}KK}}\over{16\pi}}\left(\sqrt{1-{{4m^2_{K^{\pm}}}\over{m^2}}}+\right. \right.
  \\
  & & \left. \left. \sqrt{1-{{4m^2_{K^0}}\over{m^2}}}\right)\right),
  \label{flatte}
\end{eqnarray}
where $g_{{\rm f_0}\pi\pi}=\sqrt{3/2}g_{{\rm f_0}\pi^+\pi^-}$ and 
$g_{{\rm f_0}KK}=g_{{\rm f_0}K^+K^-}=g_{{\rm f_0}K^0\overline{K^0}}$. 
These couplings have the same meaning as those defined within
  the KL frame and are related to the non dimensional
  couplings $g_{\pi}$ and $g_{K}$ (see for instance~\cite{Barberis,Aitala}) 
  through
  the relations $g_{\pi}\sim g^2_{{\rm f_0}\pi\pi}/(8\pi m^2_{\rm
    f_0})$ and $g_{K}\sim g^2_{{\rm f_0}KK}/(4\pi m^2_{\rm
    f_0})$ strictly valid only when $m\sim m_{\rm f_0}$.
In order to obtain the correct phase behaviour consistently with chiral 
perturbation theory predictions~\cite{Colangelo}, $b_0$ is expressed as
a function of the other parameters, reducing the free parameters to seven.\\

Finally in the Scattering Amplitudes (SA) fit
\cite{BP1} the amplitude is the sum of the scattering amplitudes
$T_{11}=T(\pi\pi\rightarrow\pi\pi)$ and $T_{12}=T(\pi\pi\rightarrow KK)$,
whose shapes are fixed by independent experimental information\cite{BP2}:
\begin{eqnarray}
\nonumber
A_{SA}&=&(m-m^2_0)\left(1-{{m^2}\over{s}}\right)\left[(a_1+b_1m^2+c_1m^4)T_{11}+\right.
\\
& & \left.(a_2+b_2m^2+c_2m^4)T_{12}\right]e^{i\lambda},
\label{sa}
\end{eqnarray}  
free parameters are the six
coefficients of the two polynomials, $m_0$ and the phase $\lambda$. Once the
amplitude is determined by the fit, it is analytically continued in the complex $m$ plane
and the coupling $g_{\phi}$ is determined by the pole residue. The coupling
$g_{\phi}$, having the dimension of an energy, is connected to the partial
width $\Gamma(\phi\rightarrow \gamma$f$_0$(980)) through the relation \cite{BP1}: 
\begin{equation}
\Gamma(\phi\rightarrow \gamma{\rm
  f}_0(980))={{\pi^2}\over{2}}g_{\phi}^2{{m^2_{\phi}-m^2_{\rm f_0}}\over {m^3_{\phi}}}.
\label{gphi}
\end{equation}  
\begin{figure}[ht]
  \centering
  \epsfig{file=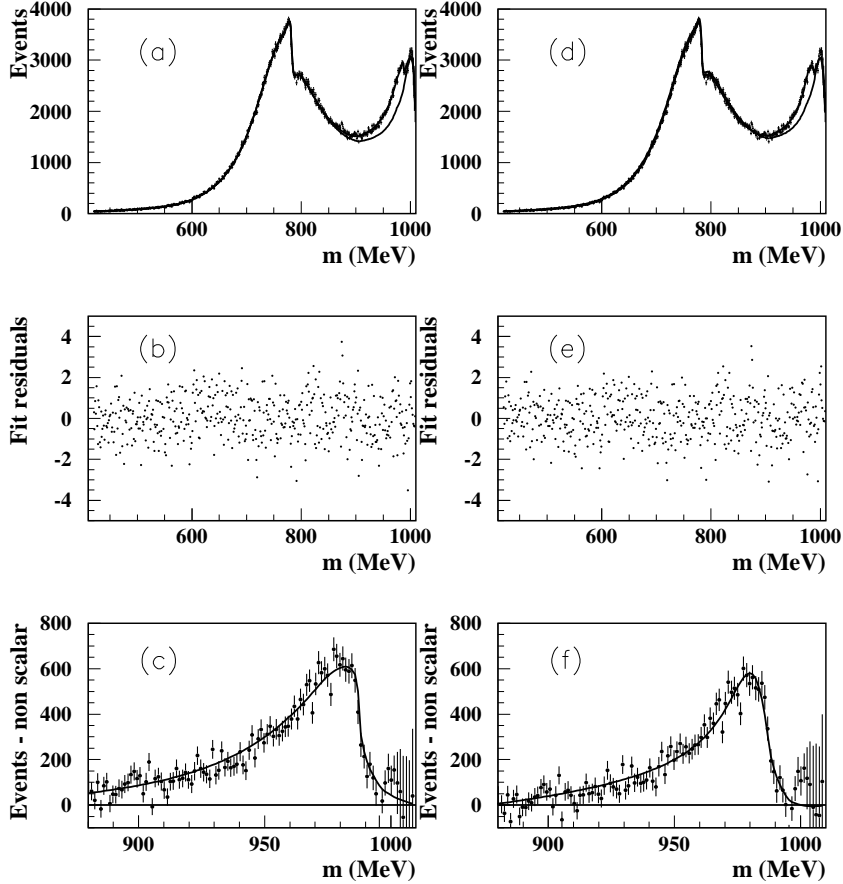,width=12cm}
  \caption{\small{\it Result of the KL fit (a)-(b)-(c) and of 
      the NS fit (d)-(e)-(f). (a)-(d) Data spectrum
      compared with the fitting function (upper curve following the data points)
      and with the estimated non-scalar part
      of the function (lower curve); (b)-(e) fit residuals as a function of $m$; 
      (c)-(f) the fitting function is compared to the spectrum
      obtained subtracting to the measured data the non-scalar 
      part of the function in the f$_0$ region. 
  }}
  \label{fitresults}
\end{figure}
\section{Results}
\label{cinque}
We fit the data in the region $420<m<1010~\mbox{MeV}$, using bins 1.2~MeV
wide \cite{nota2}.\\
First we discuss the fits KL and NS.
In both cases, a destructive interference is preferred by the fit 
for the $(d\sigma/dm)^{INT}_{\rm scal+FSR}$ term, while
the constructive interference is strongly disfavoured.
The results are shown in Fig.~\ref{fitresults}, 
the $\chi^2$ of the fits and the values of the parameters
are given in Tab.1.
The non-scalar part is well described by the parametrisation used, while we
are clearly not sensitive to the $\rho\pi$ term. 
The f$_0$ signal appears as an excess of events
in the region between 900 and 1000~MeV.
In the KL fit, the attempt to include a second scalar meson (the f$_0(600)$ or
$\sigma$), with
either E791~\cite{e791} or BES~\cite{bes} masses (respectively 478 and 541
MeV), and with free couplings, 
gives no improvement to the fit.
Moreover, since the couplings preferred by the fit
are compatible with zero, within the statistical errors,
the f$_0(600)$ is unnecessary to describe these data.\\
After the subtraction of the non-scalar part obtained in the KL and NS fits,
an asymmetric peak around 980~MeV
with a FWHM of 30$-$ 35~MeV and a height $\sim$25\% of the total is obtained,
as shown in Fig.~\ref{fitresults}(c) and (f).
Such a peak does not directly represent the f$_0$ shape but it results from the sum of a broad
term $(d\sigma/dm)_{\rm scal}$ and a negative interference term
$(d\sigma/dm)_{\rm scal+FSR}$ that cancels the low mass tail.
The NS fit requires a significantly larger value of
$\beta$ than the KL fit does. This results in  
a non-scalar part $\sim 4$\% larger in the
f$_0$ peak region, and hence a correspondingly smaller signal size.\\
\begin{table} [ht]
  \begin{center}
    \caption{\small{\it Parameter results and $\chi^2$ of the two fits  KL
        (kaon-loop) and NS
        (no-structure). The results given in parentheses
    are not directly parameters of the fits but are evaluated as functions of
    the fit parameters.
    }}
    \begin{tabular}{ccc}
      \hline
      & KL &  NS  \\
      \hline
      $\chi^2$ (p($\chi^2$))  & 538/483 (4.2\%) & 533/479 (4.4\%)\\
      \hline
      $m_{\rm f_0}$ (MeV)  & 983.0$\pm$0.6 & 977.3$\pm$0.9\\
      $g_{\phi \rm f_0\gamma}$ (GeV$^{-1}$)  &$-$ & 1.48$\pm$0.06\\
      $g_{{\rm f_0}K^+K^-}$ (GeV)  &5.89$\pm$0.14 & 1.73$\pm$0.12 \\ 
      $g_{\rm f_0\pi^+\pi^-}$ (GeV)  & (3.6) & 0.99$\pm$0.02\\ 
      $R=g^2_{{\rm f_0}K^+K^-}/g^2_{{\rm f_0}\pi^+\pi^-}$  &2.66$\pm$0.10 & (3.1)\\
      $a_0$  &$-$ &6.00$\pm$0.02\\
      $a_1$  &$-$ &4.10$\pm$0.04\\
      $b_1$(rad/GeV)  &$-$ &3.13$\pm$0.05\\
      \hline
      $m_{\rho^0}$ (MeV)  & 773.1$\pm$0.2 & 773.0$\pm$0.1\\
      $\Gamma_{\rho^0}$ (MeV)  & 144.0$\pm$0.3 & 145.1$\pm$0.1 \\
      $\alpha$ ($\times 10^{-3}$)  & 1.65$\pm$0.05 & 1.64$\pm$0.04\\
      $\beta$ ($\times 10^{-3}$)  & -123$\pm$1 & -137$\pm$1 \\
      a$_{\rho\pi}$  & 0.0$\pm$0.6 & 1.5$\pm$1.4 \\
      \hline
    \end{tabular}
  \end{center}
  \label{tab:tabresults}
\end{table}
\begin{table} [ht]
  \begin{center}
    \caption{\small{\it Intervals of maximal variations for the f$_0$ parameters
        resulting from the systematic uncertainties studies done on both fits. Notice
        that the intervals obtained are larger than the fit uncertainties given in
        the previous table.  
    }}
    \begin{tabular}{ccc}
      \hline
      parameter & KL & NS \\
      \hline
      $m_{\rm f_0}$ (MeV)  & 980$-$987 & 973$-$981\\
      $g_{\rm f_0K^+K^-}$ (GeV)  & 5.0 $-$ 6.3 & 1.6$-$ 2.3\\ 
      $g_{\rm f_0\pi^+\pi^-}$ (GeV)  & 3.0$-$ 4.2 & 0.9$-$1.1\\ 
      $R=g^2_{\rm f_0K^+K^-}/g^2_{\rm f_0\pi^+\pi^-}$  & 2.2$-$ 2.8 & 2.6$-$ 4.4\\ 
      $g_{\phi \rm f_0\gamma}$ (GeV$^{-1}$) & $-$ & 1.2$-$ 2.0\\
      \hline
    \end{tabular}
  \end{center}
  \label{tab:couplings}
\end{table}
\begin{table} [ht]
  \begin{center}
    \caption{\small{\it Results of SA fit. The $\chi^2$ and the numerical values of the
      parameters are given together with the resulting values for the
    parameters of the non-scalar part. 
    }}
    \begin{tabular}{cccc|cc}
      \hline
      $\chi^2$ (p($\chi^2$)) & 577/477 (0.1\%) &  & & & \\
\hline
      $a_1$ & 11.9 & $a_2$ & -14.7 &$m_{\rho^0}$ (MeV) & 774.4$\pm$0.2 \\
      $b_1$ & 3.3 & $b_2$ & -15.3 &$\Gamma_{\rho^0}$ (MeV) & 142.8$\pm$0.3 \\
      $c_1$ & -15.1 & $c_2$ & 35.8 &$\alpha$ ($\times 10^{-3}$) & 1.74$\pm$0.05 \\
      $m_0$ & 0. & $\lambda$ & -1.63 & $\beta$ ($\times 10^{-3}$) & -100$\pm$18 \\
        & & & & a$_{\rho\pi}$ & 0$\pm$2 \\
      \hline
    \end{tabular}
  \end{center}
  \label{tab:tsafit}
\end{table}
\begin{figure}[ht]
  \centering
  \epsfig{file=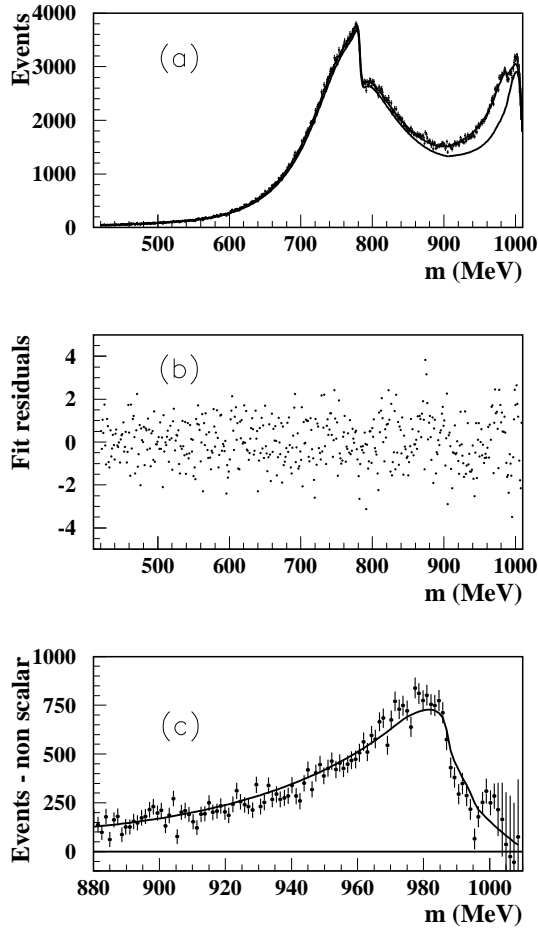,width=8cm}
  \caption{\small{\it Same as Fig.\ref{fitresults} for fit SA. In this case
  the fit lowers the non-scalar part and gives a larger signal than in the KL
  and NS fits. In the peak region
  the fit is clearly poorer than KL and NS fits.
  }}
  \label{safit}
\end{figure}
\begin{figure}[ht]
  \centering
  \epsfig{file=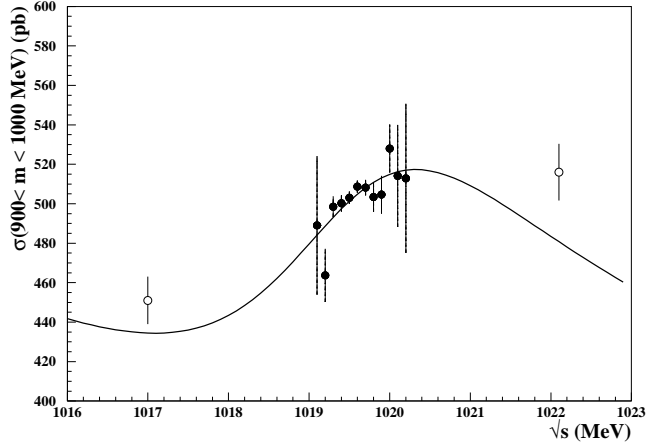,width=9cm}
  \caption{\small{\it Centre of mass energy dependence of the
      cross-section for events with $m$ in the range 900 - 1000~MeV. 
      The open points are the ``on-peak'' data sliced in 0.1~MeV wide bins, the full
      points are the ``off-peak'' data. The curve is the absolute prediction based on
      the KL fit parameters.
  }}
  \label{off-peak}
\end{figure}
\begin{figure}[ht]
  \centering
  \epsfig{file=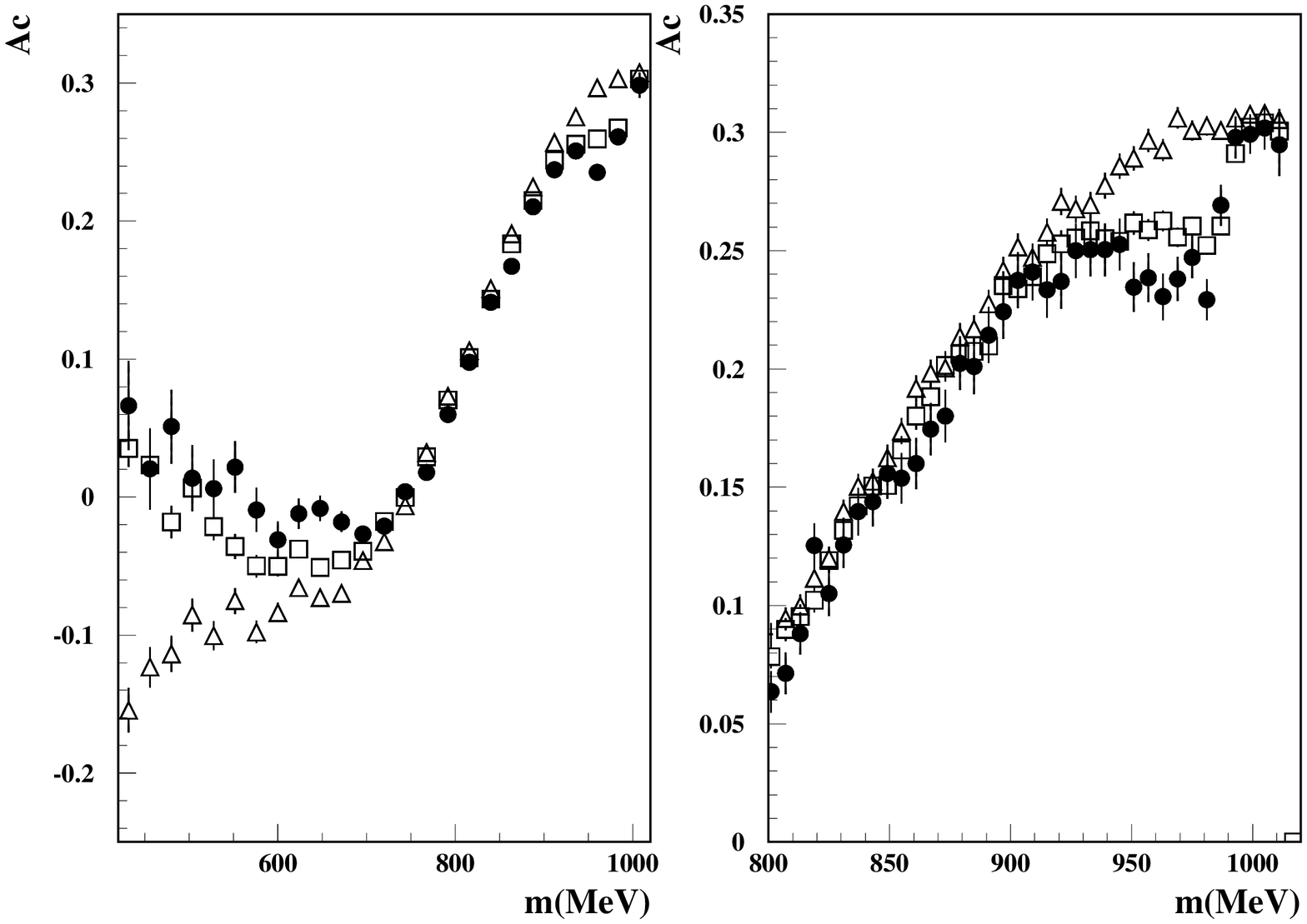,width=11cm}
  \caption{\small{\it The forward-backward asymmetry data (full circles) 
      compared to the Monte Carlo expectations based on the non-scalar part of
      the spectrum only (open triangles), and on the non-scalar plus f$_0$ part
      obtained from the KL amplitude(open squares). 
      The right plot shows the detail of the comparison in the
      f$_0$ region. 
  }}
  \label{comparison}
\end{figure}

In order to assign the uncertainty to the parameters extracted from the fits,
we have done a study of several systematic effects \cite{nota2}. We have repeated the fits
by varying the following quantities: 
the luminosity value  around its
total estimated error ($\sim 1\%$~\cite{notaDN});
the shape of the photon efficiency and the linearity of the 
photon response curves;
the size of the $\pi^+\pi^-\pi^0$ background;
the bin size, and the start and end
points of the fit. 
While repeating the fits, the parameters of the non-scalar
part are held fixed to their baseline values. 
Finally in order to take into account the systematic effect 
due to the limited knowledge of the non-scalar
part of the spectrum, the NS fit has been repeated using the non-scalar parameters
obtained from the KL fit and vice-versa.
In Tab.2 
we give the
maximal variation intervals for the parameters, resulting from the
studies discussed above.\\ 
The two fits have slightly overlapping intervals for the 
f$_0$ mass, and are both in agreement with the PDG interval 980$\pm$ 10~MeV.
We observe a large discrepancy between the KL and NS couplings 
$g_{\rm f_0\pi^+\pi^-}$ and $g_{\rm f_0K^+K^-}$. The KL fit gives couplings 
in reasonable agreement with the KLOE results obtained with the final 
state $\pi^0\pi^0\gamma$~\cite{ppg2002}.
The two fits are in agreement on the ratio 
$R=g^2_{\rm f_0K^+K^-}/g^2_{\rm f_0\pi^+\pi^-}$,
pointing to an f$_0$ more coupled to kaons than to pions.
Finally if we define an effective branching ratio
as the integral
over the full spectrum of the f$_0$ term normalised to the total $\phi$
width, we obtain 
BR($\phi\rightarrow$f$_0$(980)$\gamma$)$\times$ 
BR(f$_0$(980)$\rightarrow\pi^+\pi^-$)=   
2.1$\times 10^{-4}$ and 2.4$\times 10^{-4}$ for
KL and NS fits respectively.\\
\\  
The SA fit is shown in Fig.\ref{safit} and in Tab.3. 
The $\chi^2$ is poorer especially in
the f$_0$ peak region. Notice that a much better $\chi^2$ can be hardly
expected since $T_{11}$ and $T_{12}$ are derived from data sets that are less
accurate than the data presented here.
In any case by properly normalising the amplitude we obtain a value
$g_{\phi}\sim 6.6\times 10^{-4} {\rm GeV}$,
in agreement with the value obtained in ref.\cite{BP1} by fitting the KLOE
data on $\pi^0\pi^0\gamma$ together with other data. However we stress that 
such a value
corresponds to an effective branching ratio of $\sim 3\times 10^{-5}$, one order of magnitude lower than
the one obtained from the other two fits. 
This can be understood since in this
case the fit prefers a constructive interference term, 
hence the scalar term has a smaller
size.\\

Using the results of the KL fit,
we predict the dependence of the cross-section 
$\sigma(e^+e^-\rightarrow\pi^+\pi^-\gamma, 45^{\circ}<\theta_{\gamma}<135^{\circ}, 
900<m<1000~\mbox{MeV})$ on $\sqrt{s}$. 
The predicted behaviour is compared to the data in Fig.~\ref{off-peak},
in the $\sqrt{s}$ range between 1016 and 1023~MeV. 
Besides the ``on-peak'' data sliced in 0.1~MeV wide bins, 
we show two ``off-peak'' points taken at $\sqrt{s}$ = 1017 and 
1022~MeV, respectively. We observe a good agreement for the on-peak data,
and a marginal agreement for the two off-peak points.\\ 

Finally, following the suggestion contained in ref.~\cite{Czyc}, 
we compared the behaviour of the forward-backward asymmetry as a function
of $m$, shown in  Fig.~\ref{dati}(b), with a simulation 
including the f$_0$ contribution besides the ISR and FSR \cite{Graz}, and the effect of the $\pi^+\pi^-\pi^0$
background that dilutes the asymmetry in the low mass
region.  
The comparison is shown in Fig.~\ref{comparison}. 
The KL parametrisation of the f$_0$ amplitude has been used.
The inclusion of the f$_0\gamma$ term is essential to
have an acceptable agreement between data and simulation in the region of the
f$_0$ peak and also in the low mass region. This means that the low mass tail
of the f$_0\gamma$ amplitude that is cancelled in the $m$ spectrum by the
destructive interference with FSR, is on the contrary well evident in the $A_c$
spectrum due to the interference with ISR.
\section{Conclusion}
\label{sei}
Summarising, we found a clear evidence for the process
$\phi\rightarrow$\-f$_0$(980)$\gamma\- \rightarrow\pi^+\pi^-\gamma$ in the
$\pi\pi$ invariant mass spectrum and in the behaviour of the forward-backward
asymmetry. An
acceptable description of the data is obtained with fits KL and NS. Both fits predict the f$_0$
to be strongly coupled to kaons, the ratio $R$ between $g^2_{{\rm f_0}K^+K^-}$
and $g^2_{{\rm f_0}\pi^+\pi^-}$ being well above 2 as in the KLOE 
measurement of f$_0\rightarrow\pi^0\pi^0$ \cite{ppg2002}. The coupling to 
the $\phi$,
$g_{\phi \rm f_0\gamma}$ is found using the NS approach and is in the range
1.2 $-$ 2 GeV$^{-1}$. 
A marginal agreement is obtained by applying the fit SA.
\ack
We warmly acknowledge N.N.Achasov for many clarifications concerning the KL model;
G.Isidori, L.Maiani and S.Pacetti for the development of the
NS model here discussed and for many useful discussions;
M.E.Boglione and M.R. Pennington for providing us the scattering
amplitudes used in the fit and
O.Shekhovtsova for the asymmetry simulation.\\
We thank the DA$\Phi$NE team for their efforts in maintaining low background running 
conditions and their collaboration during all data-taking. 
We want to thank our technical staff: 
G.F.Fortugno for his dedicated work to ensure an efficient operation of 
the KLOE Computing Center; 
M.Anelli for his continuous support to the gas system and the safety of
the
detector; 
A.Balla, M.Gatta, G.Corradi and G.Papalino for the maintenance of the
electronics;
M.Santoni, G.Paoluzzi and R.Rosellini for the general support to the
detector; 
C.Piscitelli for his help during major maintenance periods.
This work was supported in part by DOE grant DE-FG-02-97ER41027; 
by EURODAPHNE, contract FMRX-CT98-0169; 
by the German Federal Ministry of Education and Research (BMBF) contract 06-KA-957; 
by Graduiertenkolleg `H.E. Phys. and Part. Astrophys.' of Deutsche Forschungsgemeinschaft,
Contract No. GK 742; 
by INTAS, contracts 96-624, 99-37; 
and by the EU Integrated Infrastructure
Initiative HadronPhysics Project under contract number
RII3-CT-2004-506078.

\end{document}